\documentclass[aps,prl,showpacs,twocolumn,superscriptaddress,amsmath,amssymb]{revtex4}
\usepackage{graphicx}
\usepackage{dcolumn}
\usepackage{bm}
\usepackage{color}
\usepackage{amssymb}
\usepackage[update,prepend]{epstopdf}
\usepackage{epstopdf}
\usepackage{ulem}
\begin{document}
\title{Measurement of the Coulomb Logarithm in a Radio-Frequency Paul Trap}

\author{Kuang Chen, Scott T. Sullivan, Wade G. Rellergert and Eric R. Hudson}
\affiliation{Department of Physics and Astronomy, University of California, Los Angeles, California 90095, USA}
 
\date{\today}

\begin{abstract}
Samples of ultracold $^{174}$Yb$^+$ ions, confined in a linear radio-frequency Paul trap, are heated via self-induced micromotion interruption, while their temperature, density, and therefore structural phase are monitored and simulated. The observed time evolution of the ion temperature is compared to a theoretical model for ion-ion heating allowing a direct measurement of the Coulomb logarithm in a linear Paul trap. This result permits a simple, yet accurate, analytical description of ion cloud thermodynamic properties, e.g. density, temperature, and structural phase, as well as suggests limits to and improvements for on-going trapped-ion quantum information efforts.
\end{abstract}
\maketitle

Binary collisions in an ensemble of charged particles are fundamental throughout physics. As such, modeling their behavior plays an important role in applications ranging from thermonuclear fusion to quantum computation. Despite their importance and the large amount of work towards understanding their effects, there is still considerable ambiguity in how to best model these collisions. Since Landau's early work~\cite{Landau1936}, the most straightforward approach handles the divergence associated with collisions of charged particles by introducing both a short-range and long-range cut-off for the $1/r$ interaction potential. The long-range cut-off is typically associated with the Debye screening length, $\lambda_D = \sqrt{\epsilon_o k_B T/\rho e^2}$, while the short-range cut-off is associated with the Landau length, $R_C = e^2/(4\pi\epsilon_o k_B T)$, which is the distance of closest approach. Taken together, the integration bounds give rise to the so-called Coulomb logarithm in its simplest form, $\ln\Lambda = \ln\left(C \lambda_D/R_C\right)$. 

Over the last 75 years, there have been many attempts to calculate an accurate form of $\ln\Lambda$, ranging from straightforward estimations of the coefficient $C$~\cite{Spitzer1962} to sophisticated analytical treatments~\cite{Brown2005} with reasonable consensus that $C \approx 0.765$~\cite{Lifshitz2006}. These results give satisfactory agreement with data for weakly coupled systems, $g = R_C/\lambda_D \ll1$, but clearly fail to describe strongly-coupled systems, $g \gg 1$, where the collision rate saturates. In this regime, more sophisticated treatments \cite{Gericke2002}, which do not require a short-range cut-off, have recommended several alternative forms for $\ln\Lambda$. And recently, a new approach, motivated by the need to model thermonuclear ignition, used molecular dynamics simulations to suggest $\ln\Lambda \sim \ln\left(1+0.7/g\right)$ \cite{Dimonte2008,Benedict2009} for $g < 10$.

Given the importance of the Coulomb logarithm, there have also been attempts \cite{Mostovych1991, Ng1995} at a direct measurement of its dependence on the strong-coupling parameter $g$. However, to date, these experiments have either been confined to the weak-coupling limit or were inconclusive because of the difficulties associated with parameterizing dense, high-energy plasmas. 

In this work, we use an alternative route to realize a strongly-coupled ion system and measure $\ln\Lambda$: laser-cooled $^{174}$Yb$^+$ ions confined in a linear Paul trap. Here, the confining trap potential provides a smoothly varying, neutralizing background for the positively charged ions, resulting in a system described as a one-component plasma. Though the density of this plasma is orders of magnitude lower than a typical strongly-coupled plasma, the low temperatures accessible through laser cooling make it possible to realize systems with $g \gg 1$. Further, by laser cooling the sample to a large $g$ and then allowing the ions to heat through self-induced micromotion interruption, we are able to measure the evolution of both the temperature and structural phase of the trapped ion cloud over a large range in $g$. From these measurements, we are then able to determine the Coulomb logarithm over a range of  $10^{-7} \leq g \leq 10^{-2}.$ Using a detailed molecular dynamics simulations, we confirm this experimental determination and extend it to a range of $10^{-7} \leq g \leq 10^{3}.$ As the values of $\ln\Lambda$ for large $g$ are known to be process dependent \cite{Drake2005}, this result must be carefully interpreted before it can be applied to other systems. Nonetheless, it offers a complete description of $\ln\Lambda$ for radio-frequency Paul traps, and thus allows a simple analytical description of trapped ion thermodynamics. 

In the remainder of this manuscript, we explain the phenomenon of ion heating by self-induced micromotion interruption, detail the method by which $\ln\Lambda$ is extracted, and describe the experimental system. We present experimental and molecular dynamics results and a recommended expression for $\ln\Lambda$ in linear Paul traps. We conclude with a discussion of the implications of this work for trapped-ion quantum information efforts.

In a linear Paul trap, an ion experiences both a time-dependent force from the confining electric potential of the trap and Coulomb repulsion from the other ions, resulting in trajectories given by:

\begin{equation}
m \frac{d^2\vec{r}_i}{dt^2} = -e\vec{\nabla}\phi(\vec{r}_i,t) + \sum_{i\neq j}^N\frac{e^2}{4\pi\epsilon_o}\frac{\vec{r}_i - \vec{r}_j}{\left|\vec{r}_i - \vec{r}_j\right|^3},
\label{eq:trajectory}
\end{equation}
with the trap potential given as:
\begin{equation}
\phi(\vec{r}_i,t) = \frac{V_{rf}}{r_o^2}(x_i^2-y_i^2)\cos(\Omega t) + \frac{\alpha V_{ec}}{z_o^2}\left(z_i^2 - \frac{1}{2}(x_i^2 + y_i^2)\right), 
\label{eq:potential}
\end{equation}
 where $r_o$ is the field radius, $z_o$ is the distance from trap center to the end cap used for axial confinement, $V_{rf}$ and $V_{ec}$ are the radio-frequency (rf) and end cap voltages, respectively,  $\Omega$ is the frequency of the rf voltage, and $\alpha$ is a geometric factor less than unity. Due to the infinite range of the Coulomb interaction, Eq. (\ref{eq:trajectory}) represents a complicated many-body problem and has no closed-form solution. Therefore, two alternative approaches are usually employed to treat this system. 
 
First, rigorous molecular dynamics (MD) simulations have been performed to study the structure of ion clouds \cite{Schiffer1988} and rf heating rates \cite{Ryjkov2005}, and for comparison with experimental ion fluorescence images \cite{Zhang2007}. Despite their successes, these simulations offer little physical intuition, making it difficult to optimize a given system. 

The second approach to modeling large ion systems in a Paul trap has been through analytical techniques \cite{Blatt1986, Siemers1988, Baba2002, Moriwaki1992}. Of the analytical approaches, the simplest and most intuitive describes the trapped ion trajectories by the well-known Mathieu solutions and includes the effect of the Coulomb interaction as hard-sphere collisions between the trapped ions to calculate, among other things,  the evolution of the trapped ion kinetic energy \cite{Baba2002, Moriwaki1992}. In this limit, the kinetic energy evolution of an ion due to the collisions with other ions is given simply as, $\dot{W} = \gamma \Delta W$, where $\gamma$ is the collision rate and $\Delta W$ is found from the kinetic energy change per collision \cite{Moriwaki1992} after enforcing conservation of momentum and energy for the collision and requiring that the new ion trajectory corresponds to a Mathieu solution. In contrast to static traps, $\Delta W$ does not average to zero over time, or over the ensemble, in a rf Paul trap. In fact,  as shown in Ref. \cite{Moriwaki1992} upon averaging $\Delta W$ is always positive, leading to the so-called micromotion-interruption heating phenomenon. Though this heating has been explained in different ways \cite{Blumel1989, Major1968, Blatt1986}, it arises from the simple fact that when ions undergo collisions their trajectories are not given by the stable Mathieu trajectories and as a result the rf trapping field can do net work on them. 

To link temperature with $W$, we introduce  $T_{sec}$ and $T_{tot}$ to be proportional to the random thermal energy (secular motion) of the ions and the total kinetic energy (secular motion plus micromotion), respectively \cite{Major1968}:
\begin{align}
  CT_{tot} = W & & CT_{sec} = W_{sec} = \eta (W-W_{ex}) 
\end{align}
where $C = \frac{3}{2}Nk_B$, $\eta$ is the ratio of secular energy to total energy and $\eta \approx \frac{3}{5}$ for low Mathieu $q$ parameter as a consequence of equipartition of energy between secular motion and micromotion \cite{Baba2002}. $W_{ex}$ accounts for the excess micromotion energy \cite{Berkeland1998} due to displacement of the ion from the node of the oscillating electric field, as a result of either the location of the ion in the crystal or stray, uncompensated, dc electric fields. Typically, $T_{ex} = W_{ex}/C$ is a few Kelvin for an ion crystal composed of $N \sim 10^3$ ions~\cite{Berkeland1998,Zhang2007}.

Using the Chandreshakar-Spitzer plasma self-collision rate~\cite{Spitzer1962}, the rate of change of the total and secular temperature of the ion cloud is given as:
\begin{align}
\dot{T}_{tot} & = \frac{e^4\rho_i(T_{sec})\ln\Lambda}{4\pi\epsilon_0^2\sqrt{m}(k_B T_{sec})^{3/2}}\overline{\epsilon}T_{tot} \nonumber \\ 
\dot{T}_{sec} & = \frac{e^4\rho_i(T_{sec})\ln\Lambda}{4\pi\epsilon_0^2\sqrt{m}(k_B T_{sec})^{3/2}}\overline{\epsilon}(T_{sec} + \eta T_{ex})
\label{eq:heating_rate}
\end{align}
where $\rho_i(T_{sec})$ is the ion density \cite{Moriwaki1992} and $\overline{\epsilon} = \Delta W / W$ is the average fractional increase of the ion energy per collision. By averaging over the rf-phase at which the collision takes place, Refs.~\cite{Baba2002, Moriwaki1992} have calculated $\overline{\epsilon} $ in terms of the Mathieu stability parameters $a$ and $q$. Through numerical integration of their result, we have found $\overline{\epsilon}$ can be simplified to $\overline{\epsilon} \approx  \frac{2}{3}(1 + 2q^{2.24})$ with a relative error $<$~0.4\% for $q\le0.4$, $a$ = 0. Thus, by laser cooling a sample of trapped ions to a low initial temperature, extinguishing the laser cooling, and monitoring the ion temperature evolution, we are able to measure $\ln\Lambda$ as a function of $g$.

The experimental apparatus used in this work consists of a sample of $^{174}$Yb$^+$ ions, loaded via laser ablation, into a linear rf Paul trap with $r_o = 1.2$~cm, $z_o = 1.075$~cm, $\eta = 0.13$, $\Omega = 2\pi\times 300$~kHz, $V_{rf}$ = 155 V and $V_{ec}$ = 5~V. A strongly coupled ion ensemble ($N = 10^{2-3}$) is realized by laser-cooling the Yb$^+$ ions, along the trap axis, with a 369~nm cooling laser (detuned from resonance by $\delta = -30$~MHz) and 935~nm repump laser ($\delta = 0$~MHz) to a starting secular temperature, measured from the Doppler broadened fluorescence profile, ranging from Doppler limited, $T_D \sim$ 1 mK to 100 mK, depending on crystal size, resulting in a one-component plasma with $g = 10^2 \sim 10^3$. 

\begin{figure}[t!]
  \centering
  \includegraphics[width=0.9\columnwidth]{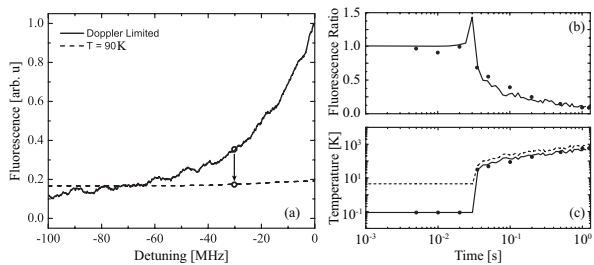}
  \caption{(a) Laser fluorescence profile for a sample of ions at the $T_D$ (solid) and at $\sim$90~K (dashed). The arrow denotes the change in fluorescence at a fixed detuning of $\delta$ = -30~MHz. (b) The observed (dots) and simulated (line) fluorescence ratio for $\delta$ = -30~MHz vs. heating time. (c) The extracted (dots) and simulated $T_{sec}$ (line) and $T_{tot}$ vs. heating time. }
  \label{fig:exp_heatingratemeas}
\end{figure}

Once the strongly-coupled plasma is established in the trap, the cooling laser is extinguished and the ions evolve in the trap and heat through self-induced micromotion heating. After a variable time delay, the cooling and repump lasers are reapplied and the fluorescence level of the ion cloud immediately recorded. If the ion temperature has increased during the time when the lasers were extinguished, this fluorescence level will be different than the steady-state value reached for the initially cold plasma, see Fig. \ref{fig:exp_heatingratemeas}(a).  By recording the ratio of fluorescence before and after heating, the temperature of the ions can be estimated as a function of heating time in a manner similar to Ref. \cite{Wesenberg2007}, as shown in Fig. \ref{fig:exp_heatingratemeas}(b)-(c) for a sample of ions with $N$ = 280. Typically, the observed fluorescence ratio decreases with increasing temperature since both the fluorescence profile is further Doppler broadened and the higher energy ion trajectories have less overlap with the laser beam inducing the fluorescence, see Fig. \ref{fig:exp_heatingratemeas}(a)-(b). However, for finite laser detuning the fluorescence ratio may actually increase for a very small temperature range just above the Doppler temperature. In this range, the fluorescence profile is slightly broadened so that the detuned laser addresses more ions. This effect gives rise to the slight peak in fluorescence ratio seen at .03~s in Fig. \ref{fig:exp_heatingratemeas}(b); similar features were observed in Ref. \cite{Wesenberg2007}.

As seen in Figs.~\ref{fig:exp_heatingratemeas}(b)-(c), the fluorescence level, and thus the temperature, is relatively unchanged until 10-100 ms after the laser cooing is extinguished, when a sharp increase in temperature occurs, followed by a region of slower heating. As detailed below, the relatively small heating  rate observed at early times is a consequence of the suppression of ion-ion collisions, i.e. a small $\ln\Lambda$, for a strongly-correlated plasma, while the sudden jump in temperature coincides with the phase transition from the liquid to gas phase. At the liquid-gas boundary, the ion density is still relatively high, but the ion motion becomes less correlated, i.e. increased $\ln\Lambda$, leading to a larger heating rate. As the ions move into the gas phase, the motion becomes even more uncorrelated, leading to a further increase in $\ln\Lambda$, however, the density, and thus the collision rate drops, leading to a reduced heating rate. Also shown in Figs.~\ref{fig:exp_heatingratemeas}(b)-(c) are the results of a molecular dynamics simulation, which initializes the ions at the experimentally realized temperature and then integrates Eq. (\ref{eq:trajectory}) numerically using a leapfrog algorithm \cite{Verlet1967} implemented in ProtoMol software \cite{Matthey2004}. As the ions heat through self-induced micromotion interruption, their fluorescence level is calculated from the known laser intensity profiles and a rate-equation model, which includes the variation of laser intensity and Doppler shift for each ion position and velocity, repectively. Given experimental imperfections, such as stray fields, machining errors, laser amplitude and frequency noise, etc., that are not included in the simulation, the agreement between the simulated and measured fluorescence ratios (Fig.~\ref{fig:exp_heatingratemeas}(b)) is satisfactory. In what follows, we use these results to extract the ion-ion heating rate and ultimately $\ln\Lambda$.  

\begin{figure}[t]
  \centering
  \includegraphics[width=0.9\columnwidth]{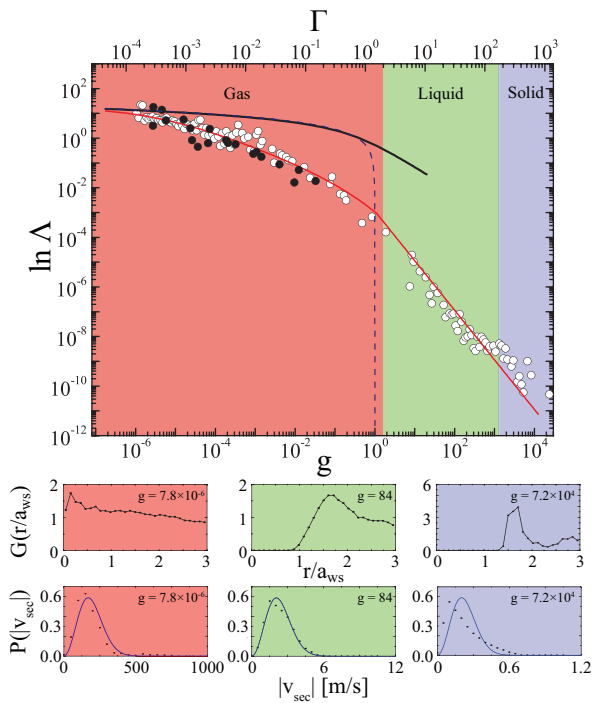}
  \caption{The experimental (black dots) and molecular dynamics (white dots) determinations of $\ln\Lambda$ versus g. Despite large variation in trap parameters (see text) the observed values fall along the same curve, indicating a `universal' form for $\ln\Lambda$. The red line represents the best fit described in the text, while the black and dashed line are the results of Ref.~\cite{Dimonte2008} and $\ln(0.765/g)$ \cite{Lifshitz2006}, respectively. Shown at the bottom of the figure are the pair correlation functions and velocity distribution functions, extracted from simulation, for selected $g$-values.}
  \label{fig:log_lambda_vs_g}
\end{figure}

Using experimental data like that shown in Fig. \ref{fig:exp_heatingratemeas} for ion clouds with $N$ between 300 and 3000 and modeling the density as 

\begin{eqnarray}
   \rho_i(T_{sec}) = \left\{
     \begin{array}{lr}
       \rho_{max} &: T_{sec} \le T_p\\
       \rho_{max}\left(\frac{T_p}{T_{sec}}\right)^{3/2} &: T_{sec} \ge T_p
     \end{array}
   \right.
   \label{eq:density}
\end{eqnarray} 
where $\rho_{max} = 4\epsilon_0 V^2_{rf}/mr_o^4\Omega^2$, $T_p = \frac{m\overline{\omega}^2}{2k_B}\left[\frac{3N}{4\pi}\frac{mr_o^4\Omega^2}{\epsilon_o V_{rf}^2} \right]^{2/3}$, and $\overline{\omega}$ is the geometric mean of the three secular frequencies, Eq. \ref{eq:heating_rate} is inverted to find $\ln \Lambda$  and the results are plotted in Fig. 2. As the heating rate is the time derivative of $T_{sec}$, the coarse granularity of the experimental data in time makes it difficult to calculate reliable values of $\ln \Lambda$ at short time scales. Therefore, we also determine the heating rate and $\ln \Lambda$ from molecular dynamics simulations. For the simulated data, the heating rate is found by taking the numerical time derivative of the ion temperature, defined by $\frac{3}{2}Nk_B T_{tot} = \frac{m}{2}\sum_{i=1}^{N}{\overline{ \vec{v}_i(t)^2 } }$ and  $\frac{3}{2}Nk_B T_{sec} = \frac{m}{2}\sum_{i=1}^{N}{\overline{ \vec{v}_{sec,i}(t)^2 } }$ where $\vec{v}_i$ and $\vec{v}_{sec,i}$ are the total and secular velocity of $i^{th}$ ion and the overline denotes averaging in principle over $t =\{-\infty,\infty\}$ -- in practice averaging over several secular motion periods is sufficient. The resulting values for $\ln\Lambda$ are consistent with those extracted from experimental data, as shown in Fig. \ref{fig:log_lambda_vs_g}, but are expected to be of higher accuracy. Using this technique, molecular dynamics simulations were performed, like those shown in Fig.~\ref{fig:exp_heatingratemeas}, for a range of ion numbers $N =\{50, 100, 500, 900\}$ and ion-cloud geometries (radial-to-axial cloud size aspect ratios \cite{Dubin1991} of  $R/z = \{ 0.25, 1, 4\}$) to determine if the parameterization of Eq.~\ref{eq:heating_rate} leads to a universal form for $\ln\Lambda$ in Paul traps.

The values of $\ln\Lambda$ extracted from the simulation are plotted versus $g$ in Fig.~\ref{fig:log_lambda_vs_g} alongside the Landau-Spitzer result \cite{Lifshitz2006} and the result of  Ref.~\cite{Dimonte2008}.  Also shown, as the top x-axis in this figure, is the corresponding plasma coupling parameter $\Gamma = e^2/\left(4\pi\epsilon_o a_{ws} k_B T_{sec}\right)$, which, given the Wigner-Seitz radius $a_{ws} = \sqrt[3]{3/(4\pi\rho_i)}$ and secular temperature, characterizes the structural phase of the ion cloud as denoted by the three regions of the graph~\cite{Brush1966}. Clearly, despite the large changes in ion number and ion-cloud geometry, the dependence of $\ln\Lambda$ on $g$ appears universal and can be parametrized by the piecewise fit:
 \begin{eqnarray}
   \ln\Lambda = \left\{
     \begin{array}{lr}
       f_{\rm{I}}(g) = \frac{\ln\left(1+0.7/g\right)}{1 + 500\sqrt{g}} & : g < 1\\
       f_{\rm{II}}(g) = \frac{f_{\rm{I}}\left(g=1\right)}{g^2} & : g \ge 1
     \end{array}
   \right.
   \label{eq:LogLambda}
\end{eqnarray} 
where the form of  $f_{\rm{I}}(g)$ has been inspired by Ref.~\cite{Dimonte2008}. Interestingly, the observed change in dependence of $\ln\Lambda$ on $g$ occurs near the gas-to-liquid phase boundary, which, since $\lambda_D/a_{ws} = 1/\sqrt[3]{3g}$, also coincides with the regime where the Debye length becomes smaller than the average inter-particle spacing. Therefore, assuming Debye theory is approximately valid for $g>1/3$, it is reasonable to expect the ion-ion cross-section is proportional to $\lambda_D^2$ and thus $\ln\Lambda \propto \lambda_D^2T^2 \propto g^{-2}$ in agreement with the fit. 

For reference, the structural phases of the ion cloud are presented in Fig. \ref{fig:log_lambda_vs_g} by the pair correlation function \cite{Brush1966} $G(r/a_{ws}) =  (4\pi\rho_i  r^2)^{-1} dN /dr$ for three different $g$ values, as determined by molecular dynamics simulation. For the solid phase $G(r/a_{ws})$ exhibits a sharp peak at $r/a_{ws} \sim 1.7$, confirming a highly-ordered crystal structure, which disappears as the ion cloud moves into the liquid and gas phases. Also, shown at the bottom of Fig. \ref{fig:log_lambda_vs_g} are the secular velocity distributions of the ions at the same three $g$ values. In this figure the points are a histogram of the simulated secular velocity distribution, while the solid curve is the Maxwell-Boltzmann (MB) velocity distribution expected for the calculated temperature. Clearly, the simulation results for the gas and liquid phases are consistent with the MB distribution, confirming the appropriateness of the Chandreshakar-Spitzer rate in deriving Eq. \ref{eq:heating_rate}. For the solid phase, the velocity distribution exhibits a significant power-law tail, violating the assumptions of Eq. \ref{eq:heating_rate} and preventing an accurate determination of $\ln\Lambda$ in this phase.

\begin{figure}[t]
  \centering
  \includegraphics[width=0.9\columnwidth]{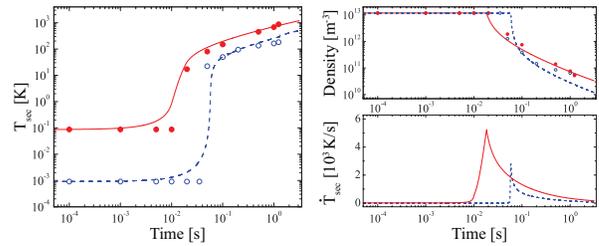}
  \caption{Comparison of experimental temperature, density and heating rate (dots) for two ion clouds ($N = 280$ (blue) and $N = 2800$ (red)) as a function of time to the result predicted by Eq. \ref{eq:heating_rate} with $\ln \Lambda$ given by Eq. \ref{eq:LogLambda} (line).}
  \label{fig:data_theory_compare}
\end{figure}

Several analytical results can be derived using Eq. \ref{eq:LogLambda} that provide insight into the plasma dynamics and have important consequences for quantum computation with trapped ions. First, for $g \ll 1$, $\ln\Lambda$ can be approximated as a constant (as is typical for low density plasma \cite{Huba1998}) and Eq. \ref{eq:heating_rate} integrated, yielding $T_{sec} \propto t^{1/3}$. In this regime, the ion temperature grows slowly until it is eventually balanced by evaporative loss from the trap or sympathetic cooling from residual neutral background gas. Second, assuming that $T_{ex} \gg T_{sec}$ in the initial ion crystal, Eq. \ref{eq:heating_rate} can be directly integrated for $g>1$ yielding $T_{sec} = [c(t_m-t)]^{-2} $, where $c = f_{\rm{I}}(1)2\pi \epsilon_0 k_B^{3/2}\overline{\epsilon}\eta T_{ex} /(\sqrt{m}e^2) $  and the time of the dramatic rise in temperature as the ions move into the gas phase is $t_m = T_{sec}^{-1/2} (t=0)/ c$. As this expression for $t_m$ ignores the contribution of $T_{sec}$ to the heating rate, it slightly overestimates the time of the phase transition; however, in the case of our experiment we have found it accurate to within $\sim$ 10 ms. Interestingly, in quantum computation with strings of trapped ions the computational gate operations occur with the laser cooling extinguished. Therefore, $t_m$ represents the upper limit for the time to implement a computational algorithm since, once the ion string melts, the quantum information is lost and the register must be reinitialized. For the parameters of \textit{e.g.} Ref. \cite{Monz2011} with 14 ions, we find this fundamental limit to be $\sim 10^3$~s. If this system is scaled to a larger number of ions, as necessary for many practical quantum computation applications, $t_m$ will be signficantly reduced if excess micromotion is not controlled and may limit the number of possible gate operations. Likewise, recent proposed experiments to use kinked ion chains to study the Kibble-Zurek mechanism \cite{DelCampo2010} and the coherence of discrete solitons \cite{Landa2010} will be fundamentally limited to timescales less than 1 $\sim$ 10 s. In addition to providing the upper limit for a single computation/simulation, the expression for $t_m$ can be used to guide future efforts. For example, linear string geometries of heavy ions at low Mathieu $q$ parameter should exhibit the longest lifetimes. 

Finally, to demonstrate the utility of the expression for $\ln\Lambda$, Fig.~\ref{fig:data_theory_compare} compares experimental data for two ion clouds of different size ($N$ = 280 and 2800), taken in the same manner as the data of Fig.~\ref{fig:exp_heatingratemeas}, with the temperature and density predicted by the integration of Eq.~\ref{eq:heating_rate} using Eq.~\ref{eq:LogLambda}. In addition to providing a simple means to accurately calculate the thermodynamical properties of a system of trapped ions,  these expressions explain several well-known experimental observations. For example, the smaller ion-ion heating rate (Fig.~\ref{fig:data_theory_compare}(c)) in the solid phase is due to ion-ion correlation as quantified through $\ln\Lambda$. 

In conclusion, we have measured the heating rate of ions trapped in a linear rf Paul trap due to self-induced micromotion interruption. These data, and detailed molecular dynamics simulations, have been used to determine the value of the Coulomb logarithm over a range of $10^{-7} \le g \le 10^3$. Though most determinations of $\ln \Lambda$ are process dependent \cite{Drake2005}, we expect our results to be comparable to $\ln \Lambda$ in other one-component plasmas for $g \ll 1$, as the Mathieu trajectories accurately describe the ion motion in this regime. This expectation is supported by the fact that our result converges to the traditional Landau-Spitzer result in this regime. However, as $g$ grows the Mathieu solutions provide a less accurate description of the ion trajectories, leading to a change in e.g. $\overline \epsilon$. If future theoretical work accounts for these effects, then our measurement might be reinterpreted to give a model independent determination of $\ln \Lambda$. Nonetheless, in its current form our result permits a simple, yet accurate, analytical description of ion cloud temperature, density, and structral phase transitions in a linear Paul trap. Thus, it should be immediately useful to a number of experimental efforts, including the growth of large ion crystals \cite{Drewsen1998}, sympathetic cooling of atomic or molecular ions \cite{Molhave2000, Blythe2005, Hudson2009}, and trapped-ion quantum information.

This work was supported by ARO grant No. W911NF-10-1-0505 and NSF grant No. PHY-1005453.

\bibliography{LogLambda}
\end{document}